\documentclass[aps,prb,twocolumn,floats,epsfig,groupaddress]{revtex4-1}
\usepackage{amssymb}
\usepackage{amsbsy}
\usepackage{amsmath}
\usepackage{xcolor}
\usepackage{bm}
\usepackage{epsfig}
\usepackage{color}
\usepackage{soul}
\usepackage{float}
\usepackage{braket}
\usepackage[colorlinks,linktocpage,bookmarks=false,citecolor=blue,linkcolor=red,urlcolor=blue]{hyperref}
\newcommand\mybar{\kern1pt\rule[-\dp\strutbox]{.8pt}{\baselineskip}\kern1pt}

\newcommand{\beq}{\begin{equation}}
\newcommand{\eeq}{\end{equation}}
\newcommand{\bea}{\begin{eqnarray}}
\newcommand{\eea}{\end{eqnarray}}

\newcommand{\SN}{S_{\rm N}}

\usepackage{xcolor}

\begin{document}
\title{Response to `Comment on Theory of growth of number entropy in disordered systems'}
\author{Roopayan Ghosh}
\author{Marko \v Znidari\v c}
\affiliation{Department of Physics, Faculty of Mathematics and Physics, University of Ljubljana, Jadranska 19, SI-1000 Ljubljana, Slovenia}

\begin{abstract}
In a recent preprint by Kiefer-Emmanouilidis, Unanyan, Fleischhauer and Sirker [arxiv:2203.06689] the authors comment on our work \cite{ghosh} which studied the number entropy in strongly disordered systems. The data presented in the comment does not refute what we have stated in our work. In fact several statements about our paper, for example, that the results we presented are due to some special initial states, are incorrect.
\end{abstract}
\maketitle

 Let us begin by summarizing the main results of our work\cite{ghosh,*PhysRevB.105.144203}. (i) We studied a disordered system at strong disorder, where we argued (supported by exact numerics) that for generic initial states; (ii) saturation value of $\SN$ does not grow with system size $L$, (iii) approach towards saturation value of $\SN$ is described well by a power law (which holds for a longer time-span than $\ln\ln t$ suggested by the previous works of the authors of the comment\cite{comm}). We also show that a simple two-state resonance model, compatible with many body localization (MBL), predicts a number of features of $\SN$, such as power law decrease of steady state $\SN$ with increasing disorder strength $W$, and the power law approach of $\SN$ to its steady state value. In short, at strong disorder we find no statistically significant signs of ergodicity.
 
In the comment, Ref.~\onlinecite{comm}, the authors present data for $\SN$ for the isotropic Heisenberg model at $W=5$, and argue that this contradicts the above findings. In what follows, we explain that this is not the case.

\textbf{(i) Disorder strength and initial states:} Firstly, in our work \cite{ghosh}, our interest was in the regime of strong disorder,  which was stated in the abstract and several places of the main text. We focused on disorder strengths $W\ge 10$, because we wanted to consider a regime where previously no ergodicity has been seen, so possibly in the MBL phase. The comment\cite{comm} mainly focuses on the disorder strength $W=5$, which is now understood to possibly be in the ergodic phase or very close to the critical point (See for example Ref.~\onlinecite{PhysRevB.98.174202} and Refs.~34,40-46 in our previous work\cite{ghosh}) and is not our regime of study. \footnote{In some cases we showed data for smaller $W$ in our work, like $W=5$, but that was used to connect our work with previous works.} Secondly, in several places of the comment\cite{comm} it has been stated that our results are due to non-generic initial states. However, this statement is not true. For our main results (ii) and (iii), we have used either averaging over random half-filled computational initial states, or even more generic half-filled initial states $|\mathcal{I}\rangle$, which is of the form $|\alpha\rangle\otimes|\beta\rangle$, $|\alpha\rangle$ and $|\beta\rangle$ being a uniform superposition of all computational states having $L/4$ particles in two subsystems of a bipartite lattice (For example in $L=4$ it is of the form $1/2[|01\rangle+|10\rangle]\otimes[|01\rangle+|10\rangle]$, see also Sec VI of our work\cite{ghosh}). Hence, in no manner, do our results depend on the choice of special initial states such as the domain-wall (DW) state. In fact, in our work\cite{ghosh} we had a separate section, Section V.C,  where we discussed the special states, the N\'{e}el and DW states. It is therefore a bit odd that while claiming our work focuses on special states, the comment\cite{comm} focuses exclusively on a particular state -- the N\'eel state. Furthermore, it is also worthwhile to remind the reader that particularly for the N\'eel state, it has already been observed that at $W=5$ there are signs of weak ergodicity \cite{PhysRevB.98.174202}.

\begin{figure}
\includegraphics[width=0.9 \columnwidth]{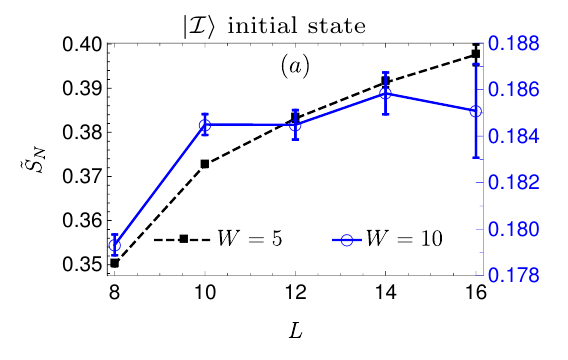}
\includegraphics[width=0.9 \columnwidth]{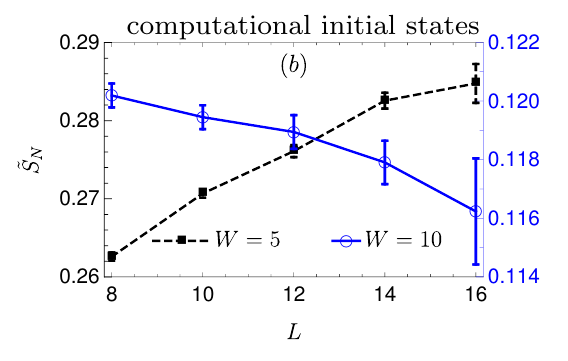}
\caption{Plots showing behaviour of $\tilde{S}_N$ for different system sizes and disorder strengths with (a) initial state $|\mathcal{I}\rangle$ and (b) random computational initial states for $W=5$  (black squares, left axis) $W=10$ (blue circles, right axis) averaged over $\sim 10^5$ configurations. \cite{Note2}}
\label{fig1}
\end{figure}
\begin{figure*}
\includegraphics[width=0.65 \columnwidth]{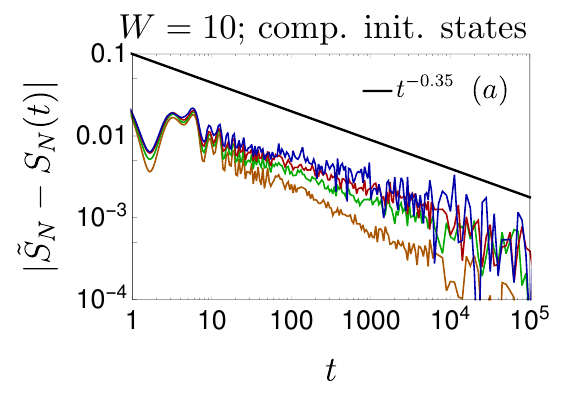}
\includegraphics[width=0.65 \columnwidth]{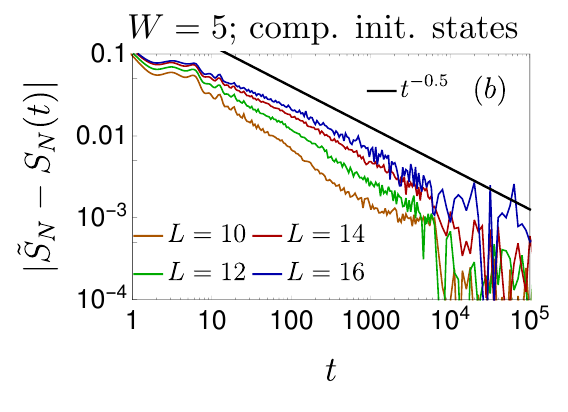}
\includegraphics[width=0.65 \columnwidth]{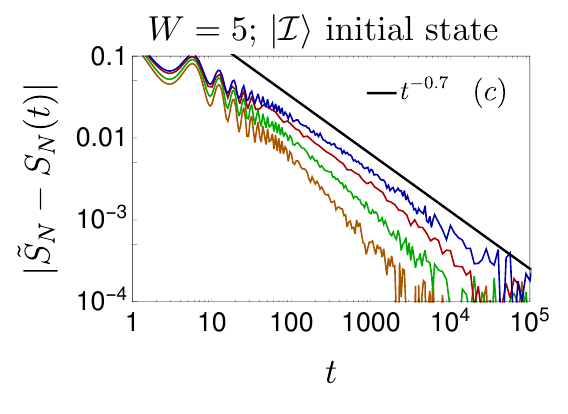}
\caption{Plots showing configuration averaged approach of number entropy to $\tilde{S}_N$ for different system sizes  and disorder strengths for (a) random computational initial states at $W=10$, (b) random computational initial states at $W=5$ and (c) initial state $|\mathcal{I}\rangle$ at $W=5$. }
\label{fig2}
\end{figure*}
\textbf{(ii) Long time saturation of $\SN$:} In Fig.~\ref{fig1}, we show the data obtained for disorder averaged long-time saturation value of $\SN$, denoted by $\tilde{S}_{\rm N}$, for the two categories of initial states considered  (similar data as our previous work\cite{ghosh}, with the addition of data for $|\mathcal{I}\rangle$ ). We clearly see in Fig.~\ref{fig1} there is no statistically significant increase of $\tilde{S}_N$ with system size at  large disorder, $W=10$, for both computational and more generic $|\mathcal{I}\rangle$ initial states, hinting at localization rather than ergodicity. For $W=5$ there is indeed an increase of $\tilde{S}_N$ seen in both cases with $L$, as pointed out in the comment\cite{comm} and already in our previous work\cite{ghosh}, but this is known to be a regime of weak ergodicity\cite{PhysRevB.98.174202}. Furthermore, as we also showed in our previous work, the saturation values are far from being ergodic. The error bars are also significant even with a sample size of $10^5$, hence one needs to be very careful while attempting any fits to such data.

\textbf{(iii) Power law approach of $\SN$ to saturation:} In the comment\cite{comm}, the authors show how fitting a power law growth of the form $A-B/t^{\alpha}$ with $\alpha=1$ does not agree with the data.  This observation is not new, since we had already shown in Fig. 14(b) of our work\cite{ghosh} that at $W=10$, $\alpha$ is $0.7$ and not $1$ for $|\mathcal{I}\rangle$ initial state, and is independent of system size, along with an explanation of the limitations of the two-state model. In Fig.~\ref{fig2}(a) we show a similar plot for computational states, where again for $L\sim 12-16$, there is no statistically significant drift of $\alpha$ with $L$. Even for lower disorder $W=5$, for the two types of initial states, the tails at long time ($t \sim 100-10^4$), which is the quantity of interest, show a similar exponent at different system sizes as evident from Figs.~\ref{fig2}(b) and (c).

\begin{figure}[H]
\includegraphics[width=0.8 \columnwidth]{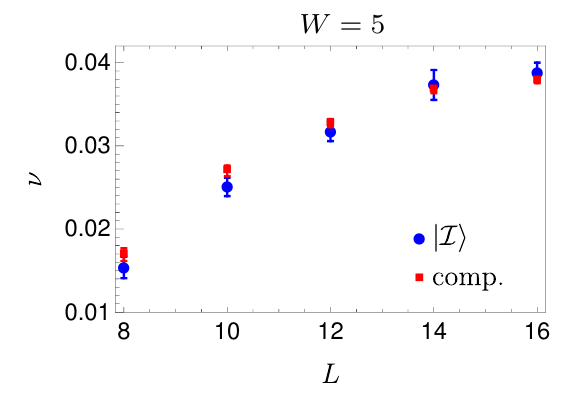}
\caption{Plot of $\nu$ obtained from fitting configuration averaged $\SN(t) = \mu +\nu \ln \ln t$ for $|\mathcal{I}\rangle$ and computational (comp.) initial states vs $L$ at $W=5$.}
\label{fig3}
\end{figure}

 On the other hand, the $\nu \ln \ln t$ fit proposed by the authors of the comment\cite{comm} is not foolproof unlike the claim. It is valid for a very small window of $t$, the value of $\nu$ actually shows a drift with $L$ for the optimal choice of fit window, and has $\chi^2$ per data point at least twice as large as power law fits. Picking an optimal window for the fit, in Fig.~\ref{fig3} we show the increase in $\nu$ with $L$ at $W=5$ for the two sets of initial states considered here. One can see from Fig.~\ref{fig3} that,  from $L\sim 12-16$, where $\ln L$ increases by $11\%$ (since the relevant scale is $\ln L$), $\nu$ increases by $25\%$, which is significant. To obtain the optimal window for this fit, we have chosen a window of time $t_1$-$t_2$, that minimizes $\chi^2$ per data point. $t_1$ is chosen approximately at $t_1 \sim 5$ which is similar to the time chosen in previous works of the authors of the comment\cite{comm}, and on minimizing $\chi^2$, $t_2$ is obtained as $\sim 100-300$ for computational states and $\sim 300-500$ for $|\mathcal{I}\rangle$ states. Unlike such a small window fit of $\ln \ln t$ with $L$-dependent $\nu$, our power law fit with almost constant $\alpha$ is asymptotic, and therefore holds for arbitrarily large $t_2$ (in practice upper bounded by fluctuations due to finite sample). One can see this from Fig,~\ref{fig2}, where the power law fits remain valid at least till $t\sim 10^4$. It is also worthy to note that on hand-picking the window by changing $t_2$, it is in principle possible to make $\nu$ constant across system sizes, but this  has an adverse effect on the $\chi^2$ of the fit, which increases by several times.
 
 To summarize, in Ref.~\onlinecite{ghosh}, we explained the slow growth of $\SN$ at strong disorder via many-body resonances, which we modeled via a two-state model to predict several results, such as the $1/W$ ($1/W^2$) decrease of mean (median) long time saturation $\SN$ and power law approach of $\SN$ to its saturation, which qualitatively agreed with exact numerics. We concluded there was no need to invoke eventual ergodicity to explain the observed growth. We never claimed that the two-state model is perfect and had already shown its limitations in predicting the accurate exponent in our work. On the other hand the comment\cite{comm} is based solely on fitting which is done over a small window without a precise theoretical explanation. Also, the $\alpha\rightarrow 0$ limit of the Renyi number entropy, $S^{(\alpha)}_{\rm N}$, the Hartley entropy, studied in Fig. 3 of the comment\cite{comm} seems to be a hand-picked quantity to support the idea of slow particle transport in such systems. Because it essentially measures logarithm of the number of occupied number-blocks, it is likely even more prone to finite size effects than $\SN$, in addition to being plagued by the problem of over-fitting in a narrow window.
 
\textbf{Conclusion:}  Ultimately, without access to large system sizes one cannot say with certainty what is the correct physics in the thermodynamic limit. The verifications discussed here rely on small-$L$ numerics (both our data and their data use small system sizes), hence, we will refrain from making categorical judgments like Ref.~\onlinecite{comm} do for example in their last sentence, ``To summarize, the theory [1]($A+B/t$ \textit {fit}) does not describe the data. They are instead well described by [2]($\nu \ln \ln t$ \textit {fit}).", and rather leave readers to objectively decide which is the simpler and more accurate explanation compatible with the data. We prefer the simpler one\cite{Mazin2022} even if it seems trivial to some.

\begin{thebibliography}{5}%
\makeatletter
\providecommand \@ifxundefined [1]{%
 \@ifx{#1\undefined}
}%
\providecommand \@ifnum [1]{%
 \ifnum #1\expandafter \@firstoftwo
 \else \expandafter \@secondoftwo
 \fi
}%
\providecommand \@ifx [1]{%
 \ifx #1\expandafter \@firstoftwo
 \else \expandafter \@secondoftwo
 \fi
}%
\providecommand \natexlab [1]{#1}%
\providecommand \enquote  [1]{``#1''}%
\providecommand \bibnamefont  [1]{#1}%
\providecommand \bibfnamefont [1]{#1}%
\providecommand \citenamefont [1]{#1}%
\providecommand \href@noop [0]{\@secondoftwo}%
\providecommand \href [0]{\begingroup \@sanitize@url \@href}%
\providecommand \@href[1]{\@@startlink{#1}\@@href}%
\providecommand \@@href[1]{\endgroup#1\@@endlink}%
\providecommand \@sanitize@url [0]{\catcode `\\12\catcode `\$12\catcode
  `\&12\catcode `\#12\catcode `\^12\catcode `\_12\catcode `\%12\relax}%
\providecommand \@@startlink[1]{}%
\providecommand \@@endlink[0]{}%
\providecommand \url  [0]{\begingroup\@sanitize@url \@url }%
\providecommand \@url [1]{\endgroup\@href {#1}{\urlprefix }}%
\providecommand \urlprefix  [0]{URL }%
\providecommand \Eprint [0]{\href }%
\providecommand \doibase [0]{http://dx.doi.org/}%
\providecommand \selectlanguage [0]{\@gobble}%
\providecommand \bibinfo  [0]{\@secondoftwo}%
\providecommand \bibfield  [0]{\@secondoftwo}%
\providecommand \translation [1]{[#1]}%
\providecommand \BibitemOpen [0]{}%
\providecommand \bibitemStop [0]{}%
\providecommand \bibitemNoStop [0]{.\EOS\space}%
\providecommand \EOS [0]{\spacefactor3000\relax}%
\providecommand \BibitemShut  [1]{\csname bibitem#1\endcsname}%
\let\auto@bib@innerbib\@empty
\bibitem [{\citenamefont {Ghosh}\ and\ \citenamefont {\v{Z}nidari\v{c}}(2021)}]{ghosh}%
  \BibitemOpen
  \bibfield  {author} {\bibinfo {author} {\bibfnamefont {R.}~\bibnamefont
  {Ghosh}}\ and\ \bibinfo {author} {\bibfnamefont {M.}~\bibnamefont
  {\v{Z}nidari\v{c}}},\ }\href {https://arxiv.org/abs/2112.12987} {\bibfield  {journal}
  {\bibinfo  {journal} {arXiv:2112.12987v2}\ } (\bibinfo {year}
  {2021})}\BibitemShut {NoStop}%
\bibitem [{\citenamefont {Ghosh}\ and\ \citenamefont {\ifmmode \check{Z}\else
  \v{Z}\fi{}nidari\ifmmode~\check{c}\else
  \v{c}\fi{}}(2022)}]{PhysRevB.105.144203}%
  \BibitemOpen
  \ \bibinfo {note} {published }\href {\doibase 10.1103/PhysRevB.105.144203} {\bibfield  {journal} {\bibinfo
  {journal} {Phys. Rev. B}\ }\textbf {\bibinfo {volume} {105}},\ \bibinfo
  {pages} {144203} (\bibinfo {year} {2022})}\BibitemShut {NoStop}%
\bibitem [{\citenamefont {Kiefer-Emmanouilidis}\ \emph
  {et~al.}(2022)\citenamefont {Kiefer-Emmanouilidis}, \citenamefont {Unanyan},
  \citenamefont {Fleischhauer},\ and\ \citenamefont {Sirker}}]{comm}%
  \BibitemOpen
  \bibfield  {author} {\bibinfo {author} {\bibfnamefont {M.}~\bibnamefont
  {Kiefer-Emmanouilidis}}, \bibinfo {author} {\bibfnamefont {R.}~\bibnamefont
  {Unanyan}}, \bibinfo {author} {\bibfnamefont {M.}~\bibnamefont
  {Fleischhauer}}, \ and\ \bibinfo {author} {\bibfnamefont {J.}~\bibnamefont
  {Sirker}},\ }\href {https://arxiv.org/abs/2203.06689} {\bibfield  {journal}
  {\bibinfo  {journal} {arXiv:2203.06689}\ } (\bibinfo {year}
  {2022})}\BibitemShut {NoStop}%
\bibitem [{\citenamefont {Doggen}\ \emph {et~al.}(2018)\citenamefont {Doggen},
  \citenamefont {Schindler}, \citenamefont {Tikhonov}, \citenamefont {Mirlin},
  \citenamefont {Neupert}, \citenamefont {Polyakov},\ and\ \citenamefont
  {Gornyi}}]{PhysRevB.98.174202}%
  \BibitemOpen
  \bibfield  {author} {\bibinfo {author} {\bibfnamefont {E.~V.~H.}\
  \bibnamefont {Doggen}}, \bibinfo {author} {\bibfnamefont {F.}~\bibnamefont
  {Schindler}}, \bibinfo {author} {\bibfnamefont {K.~S.}\ \bibnamefont
  {Tikhonov}}, \bibinfo {author} {\bibfnamefont {A.~D.}\ \bibnamefont
  {Mirlin}}, \bibinfo {author} {\bibfnamefont {T.}~\bibnamefont {Neupert}},
  \bibinfo {author} {\bibfnamefont {D.~G.}\ \bibnamefont {Polyakov}}, \ and\
  \bibinfo {author} {\bibfnamefont {I.~V.}\ \bibnamefont {Gornyi}},\ }\href
  {\doibase 10.1103/PhysRevB.98.174202} {\bibfield  {journal} {\bibinfo
  {journal} {Phys. Rev. B}\ }\textbf {\bibinfo {volume} {98}},\ \bibinfo
  {pages} {174202} (\bibinfo {year} {2018})}\BibitemShut {NoStop}%
\bibitem [{Note1()}]{Note1}%
  \BibitemOpen
  \bibinfo {note} {In some cases we showed data for smaller $W$ in our work,
  like $W=5$, but that was used to connect our work with previous
  works.}\BibitemShut {Stop}%
\bibitem [{Note2()}]{Note2}%
  \BibitemOpen
  \bibinfo {note} {In all the figures, for initial state  $|\mathcal{I}\rangle$, a configuration means a disorder realization, whereas for the computational state it means a randomly chosen initial state from the computational basis states and a random disorder realization.}\BibitemShut {Stop}%
\bibitem [{\citenamefont {Mazin}(2022)}]{Mazin2022}%
  \BibitemOpen
  \bibfield  {author} {\bibinfo {author} {\bibfnamefont {I.}~\bibnamefont
  {Mazin}},\ }\href {\doibase 10.1038/s41567-022-01575-2} {\bibfield  {journal}
  {\bibinfo  {journal} {Nature Physics}\ }\textbf {\bibinfo {volume} {18}},\
  \bibinfo {pages} {367} (\bibinfo {year} {2022})}\BibitemShut {NoStop}%

\end{thebibliography}
\end{document}